\documentclass[11pt,a4paper]{article}
\pdfoutput=1
\usepackage{jcappub}
\usepackage{verbatim}
\usepackage{amsmath}
\usepackage{indentfirst}
\usepackage{mwe}    % loads »blindtext« and »graphicx«
\usepackage{subfig}
\usepackage{slashed}
\usepackage{tensor}
\usepackage{tikz}
\usetikzlibrary{decorations.markings,decorations.pathmorphing}
\usetikzlibrary{patterns}
\usepackage{comment,hyperref,float}

\title{Observational constraints on Hyperinflation}
\author{Marios Bounakis,}
\emailAdd{m.bounakis2@ncl.ac.uk}
\author{Ian G. Moss}
\emailAdd{ian.moss@ncl.ac.uk}
\author{and Gerasimos Rigopoulos}
\emailAdd{gerasimos.rigopoulos@ncl.ac.uk}
\affiliation{School of Mathematics, Statistics and Physics, Herschel Building, Newcastle University, Newcastle upon Tyne, NE1 7RU, UK}

\abstract{ We study a Hyperinflation model involving a field doublet on a hyperbolic field-space manifold and an exponential potential, providing a concise treatment of the evolution of the entropic and adiabatic perturbations around the  homogeneous hyperbolic attractor solution. We find that the adiabatic spectral index narrows down the admissible values of the potential's slope to a very small region, severely restricting the state space of the allowed background solutions.}

\begin{document}

\maketitle

\section{Introduction}
Since its inception, Inflation \cite{Guth:1980zm} has been extensively tested both theoretically and observationally. The model's striking successes, but also its shortcomings, have resulted in a multitude of subsequent developments in model building, starting from single-field slow-roll inflation \cite{Linde:1981mu} (For an excellent review, see \cite{Baumann:2009ds} ) to more complex multi-field models \cite{Easson:2007dh,Cremonini:2010ua,Achucarro:2015rfa,Achucarro:2016fby}  (For a broad review on multi-field inflation, see \cite{Gong:2016qmq}). One of the most sought after objectives has been the embedding of an inflationary scenario as an effective low energy limit in a more fundamental high energy theory - such as String Theory \cite{Baumann:2014nda}. Recent investigations have resulted in deSitter space being excluded as a string theory vacuum \cite{Obied:2018sgi}, placing simple single-field slow-roll inflationary models in the Swampland \cite{Garg:2018reu,	Vafa:2005ui,Ooguri:2006in}, with the introduction of the corresponding Conjectures \cite{Agrawal:2018own,Obied:2018sgi}. These considerations indicated that the Swampland constraints would favour multi-field inflationary models with potentially non-trivial kinetic terms \cite{Berglund,Achucarro:2018vey}. 
	
A model dubbed Hyperinflation \cite{Brown} has become topical in recent years due to its capacity to evade the Swampland while satisfying observational results. In its introduction, Brown presents the model and its connection to Spinflation \cite{Easson:2007dh}, presenting how Hyperinflation can produce a large number of e-folds, even for steep potentials, therefore relaxing the constraint for a slow-roll regime, and seed our Universe's structure through adiabatic perturbations; all the while motivating the possible existence of Hyperinflation signatures in the equilateral non-Gaussianity. These qualitative statements were detailed quantitatively in \cite{Mizuno}, and both the background dynamics and the dynamics of the perturbations of the fields were analysed resulting in analytical expressions for cosmological parameters (power spectrum of curvature perturbation and its spectral index) being presented in the slow-varying approximation. More recently, Hyperinflation was generalised to models with an arbitrary number of fields and for various potentials, even non-rorationally symmetric ones \cite{Bjorkmo}, and was presented as being able to lay the groundwork for observationally viable models that satisfy the Swampland Conjectures \cite{Cicoli:2020cfj}. Lastly, the original indication by Brown in \cite{Brown} that Hyperinflation is an attractor in the regime that slow-roll is a repeller was extensively studied in the context of dynamical attractors \cite{Renaux-Petel:2015mga,Cicoli:2018ccr,Cicoli:2019ulk, Christodoulidis}. 
	
In this work, we draw inspiration from the aforementioned works and thoroughly examine linearized perturbations in a model with a 2-dimensional hyperbolic field space and an exponential potential that admits a scaling solution. We start by reviewing the background dynamics and manifestly showcase the parameters for which the gradient and Hyperinflation solutions are stable (unstable). We then obtain the evolution for the fields' linear perturbations and numerically evaluate the power spectrum and the spectral index for a range of values of the constant normalised Killing direction velocity, ${y}$, of the hyperinflation solution. We find that observational results place tight bounds on its permissible values within the narrow region of $0.74 \times\, 10^{-2}\leq {y}\leq 0.94 \,\times \,  10^{-2}$ and  $0.1603 \leq {y}\leq 0.1774$, corresponding to very a very narrow range of admissible potential slopes  $0.54 \times 10^{-4}\leq p-3\leq0.88\times 10^{-4} $ and $0.025\leq p-3\leq 0.031$. Ultimately, working in the small $\epsilon$ regime we find there is a maximal value for the $\epsilon$ parameter allowed by observations.

\section{Model and Background Field Considerations}
	
\subsection{Fundamentals}
	
The action for the model we consider involves a scalar doublet with a hyperbolic $\mathbb{H}^2$ field space geometry, minimally coupled to gravity and normalised by the $\mathbb{H}^2$ field space scale $M_H$, 
$\varphi^a= \phi^a/M_H$, and is given by: 
\begin{equation}
\label{action}
S=\frac12
\int\, d^Dx \,  \sqrt{-g}  \, \left[M_p^2 \, R- M_H^2 \mathcal{G}_{ab}(\varphi) \partial_{\mu}\varphi^a\partial_{\nu}\varphi^b\, g^{\mu\nu} 
- 2M_H^2 V \right],
\end{equation}
where $V(\varphi^a)$ is the scalar field potential. In this work, we adopt the notation that latin letters correspond to field space indices $\{a,b,..\}=\{1,2\}$, related to the field space metric $\mathcal{G}^{ab}$, while Greek letters $\{\mu,\nu,..\}=\{0..4\}$ correspond to spacetime indices and are associated to a Friedmann-Robertson-Walker (FRW) spacetime metric
\begin{equation}
ds^2=g^{\mu\nu}dx^{\mu} dx^{\nu}=-dt^2+a(t)\delta_{ij}dx^i dx^j
\end{equation} 
describing a homogeneous and isotropic Universe. As usual, the Hubble parameter is defined as $H=\partial_t a/a$.  

Varying the action with respect to the doublet and the metric leads to the field equations of motion and the Einstein equation:
\begin{equation}
\begin{split}
\label{background_eom}
&\mathcal{D}^{\mu}\mathcal{D}_{\mu} \, \varphi^a - \mathcal{G}^{ab} \, V,_b=0 \\
&M_p^2 \, G_{\rho \sigma}  + M_H^2 \left(
\frac12 g_{\rho \sigma} \, \mathcal{G}_{ab} \, \varphi^a,_{\mu} \, \varphi^b,^{\mu} + \, g_{\rho\sigma} \, V - 
 \, \mathcal{G}_{ab}\, \varphi^a,_{\rho} \, \varphi^b,_{\sigma}\right)=0.
\end{split}
\end{equation}
As usual $G_{\mu\nu}$ is the Einstein tensor and we have introduced the covariant derivative
\begin{equation}
\label{covariant_derivative}
	\mathcal{D}_{\mu} V^a = \partial_{\mu}V^a + \Gamma^a_{bc} \, \partial_{\mu} \, \varphi^b \,  V^c
\end{equation} 
associated with the spacetime metric $g_{\mu\, \nu}$ acting on a field space vector $V^a\, [\varphi^b(x^{\mu})]$ through the pushforward operator $\left[\mathcal{O}_{\star}\right]^a_{\ \mu}= \varphi^a,_{\mu}$ which permits the association of tensors on the spacetime and the field-space manifold, $D^a=\varphi^a{}_{,\mu} \mathcal{D}^{\mu}$.
When homogeneity and isotropy are imposed, equations \eqref{background_eom} reduce to:
\begin{align}
\label{simple_background_eom}
&\mathcal{D}_t\partial_t \varphi^a + 3\, H\, \partial_t\varphi^a+ \mathcal{G}^{ab} \, V,_b=0,\\
&3H^2=\kappa^2\left(\frac12\mathcal{G}_{ab} \, \partial_t\varphi^a \partial_t\varphi^b+V\right),\\
&\partial_t H=-\frac12\kappa^2\mathcal{G}_{ab} \, \partial_t\varphi^a \partial_t\varphi^b.
\end{align}   
where
\begin{equation}
\kappa^2=\frac{M_H^2}{M_p^2}.
\end{equation}

%-----
%Recent work has motivated the consideration of multifield dynamics in an Inflationary setting.
%
%\textbf{
%Link between jordan frame non-minimal couplings and the conformal transformation to the einstein frame. Link to Grav.Corrections and VdW action? }
%
%\textbf{Review of sfakianakis' background- Spiral p.9-10 short discussion} \cite{Christodoulidis} \cite{Garcia-Saenz} \cite{Brown} \cite{Mizuno}

Despite the fact that higher dimensionality field-spaces ($d  >2$ ) can give rise to a plethora of interesting realisations (and have been argued to even take inflation out of the Swampland \cite{Bjorkmo}), we choose to restrict ourselves to a 2-dimensional field-space manifold, as it proves to be a sufficiently illustrative example of the intriguing effects that the non-trivial field-space geometry gives rise to. Focusing on the hyperbolic inflation scenario \cite{Christodoulidis,Brown,Mizuno} in which the space has a transitive isometry (in the sense of shifts in the $\varphi^2$ direction), leads to a metric of the form 
\begin{equation}
\label{fieldmetric}
\mathcal{G}_{ab}=\begin{bmatrix} 
1 & 0  \\
0 & f^2(\rho) \\
\end{bmatrix}
\end{equation}
for the field directions $\varphi^1\equiv \rho $, $\varphi^2\equiv\varphi $.
One parametrisation of the hyperbolic field space is the Poincare half-plane defined by $f(\rho)= \ \sinh({\rho})$. For large radial-direction field values, $\rho \gg 1 $, %the $G_{22}$ component of the metric \eqref{fieldmetric} takes the form  
the field-space geometric function $f(\rho)= \sinh(\rho)\approx \,e^{\rho}$.

For the metric \eqref{fieldmetric}, the only non-vanishing Christoffel symbols and the Ricci scalar (relevant for the next section), prove to be
\begin{equation}
\label{geometry}
\Gamma^{\rho}_{\phi \, \phi}=-f^2(\rho), \ \ \ \ \ \Gamma^{\phi}_{\rho \, \phi }={1}, \ \ \ \ \
\mathcal{R}=-2\frac{\partial^2_{\rho}f(\rho)}{f(\rho)}.
\end{equation}

Let us assume an exponential potential that also respects this isometry of the field space and depends only on the ``radial'' field-space direction:
\begin{equation}
\label{potential}
V(\rho)\approx V_0\, e^{\kappa^2\, p\, \rho },
\end{equation}
where $p$ is a positive constant. The potential gradient in the $a$ direction is denoted by $p_a$,
\begin{equation}
p_a=  \kappa^{-2} \, \frac{\partial(\ln V)}{  \partial \phi^a }.
\end{equation}
Following the work of \cite{Christodoulidis}, the system of Klein-Gordon equations \eqref{simple_background_eom}
can be transformed into an autonomous system for the normalised field space coordinates $\varphi^a$ and their velocities 
$v^a=d\varphi^a/dN = \dot{\varphi}^a/H \equiv {\varphi^a}'$, with respect to the e-fold number $dN=H\ dt$,
\begin{equation}
  \begin{split}
  \label{autosystem}
   {\varphi^a}^{\prime} =&v^a \\
   {v^a}^{\prime}    =&-(3-\epsilon )(v^a+p^a)-\Gamma^a_{bc} \ v^b \ v^c.
  \end{split}
\end{equation}
Here, $\epsilon=-\dot{H}/H^2$ is the rate of change of the Hubble parameter which, using \eqref{simple_background_eom}, can be written as 
\begin{equation}
\label{epsilon}
	\epsilon = \frac12\kappa^2 \, v_a \, v^a.
\end{equation}
and its evolution equation is given by 
\begin{equation}
\label{epsilonprime}
\epsilon^{\prime}=-(3-\epsilon)(2\kappa^{-2}\epsilon-p_a\, v^a)
\end{equation}

%The Christoffel connections associated with this metric 
It is useful, here, to define the coordinates
$(x,y)\equiv(v^{\rho}, f(\rho) \, v^{\phi})$, namely, the projections of the field-space velocity orthogonal to and along the Killing direction $k_a=(0,f(\rho))$. In these coordinates, the autonomous system \eqref{autosystem} becomes:
\begin{equation}
\begin{split}
 {x}' = &-(3-\epsilon)({x}+ {p}_{\rho}) + {y}^2\\
 { y }'= &-(3-\epsilon+{x} )\, { y }
\end{split}
\end{equation}
and \eqref{epsilon} takes the useful form
\begin{equation}
\label{epsilonxy}
\epsilon = \frac12\kappa^2 \, [x^2+y^2].
\end{equation}

\subsection{Background field and linear stability in deSitter}

%Taking the quasi-classical limit $M_p^2 \rightarrow \infty$ 

%For the rest of this work, we restrict ourselves to solutions relating to deSitter space.  

\begin{figure}[H]
	\centering	
	\begin{minipage}{0.8\textwidth}
		\centering
		\includegraphics[width=0.65\linewidth]{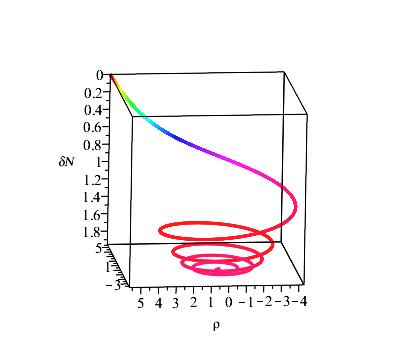}
		\caption{ Evolution into the homogeneous hyperbolic solution for ${ p}=3.01$, with the vertical axis denoting the elapsed number of e-folds from the beginning of the trajectory.\label{spiral} }
	\end{minipage}\hspace{0.5cm}
\end{figure}

Scaling solutions by definition satisfy $\epsilon'=0$, hence, from \eqref{epsilonprime}
\begin{equation}
	\epsilon=- \frac12\kappa^2 p_a\, v^a \, .
\end{equation}
The system \eqref{autosystem} results in two types of scaling solutions:
\begin{itemize}
	\item The gradient solution: $y=0$. In this case, ${x}=-{ p_{\rho}}$. To examine the stability of this solution we obtain the local Lyapunov exponents (the eigenvalues of the matrix $J$ for the linearised system: $({x}^a)'=J^a_{\ b } \ {x}^b$) which are $(\lambda_1,\lambda_2)=(-3, p-3) $. The solutions evolve towards the gradient solution as: 
\begin{equation}
\begin{split}
{\rho}={\rho}(0)-\frac13 { u }_{\rho}(0)\, e^{-3N}-{ p_{\rho} }N \\
{\phi}={ \phi}(0)+\frac{1}{p_{\rho}-3} u_{\phi}(0)\, e^{(p_{\rho} -3)N}
\end{split}
\end{equation}	
	For $p<3$, the angular velocity vanishes exponentially fast irrespective of any initial value, while the radial veloxity approaches the critical point value ${ x }=-{ p_{\rho} }$ and therefore this solution is stable. This can be explicitly seen from the phase space diagram, in figure\ref{fig:background}. 
	%As it can be seen in figure\ref{gradplot}, for small values of $ {p}<3$, the gradient solution is stable and the system evolves along the direction orthogonal to the Killing. 
For steeper potentials, $p>3$ the gradient solution (dashed black, red and purple lines in figure\ref{fig:background}) are unstable and for any initial non-trivial value of ${y}$, the system diverges from the gradient solution critical point $(x,y)=(-p,0)$ and evolves exponentially fast towards the hyperbolic solution critical point. 

\begin{figure}[H]	
	\includegraphics[width=0.45\linewidth]{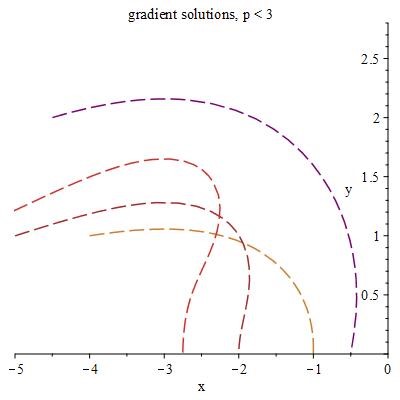}
	\includegraphics[width=0.45\linewidth]{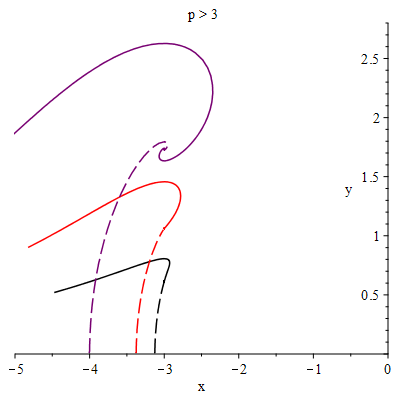}
	\caption{Phase diagram of $[ {x}(t), {y}(t)]$, with initial conditions away from the critical point, for various values of the potential gradient ${p}$. For $p<3$ the system evolves to the gradient solution, while for $p\geq3$ the system will evolve towards the hyperinflation solution, irrespective of the initial conditions. On the right panel, dashed lines commence with zero angular velocity ($y=0$) but evolve to the Hyperbolic solution. The red solid line corresponds to a potential with $p=3$.}
	\label{fig:background}
\end{figure}

\item The hyperbolic solution: Here, the system is allowed to assume velocity in both directions (see figure\ref{spiral}) and consequently, evolves towards the Hyperinflation critical point 
\begin{equation}
\label{hypercrit}
({ x }  , { y } )_{Hyper}=\left(-3, \pm \sqrt{3{ p }-9} \right).
\end{equation} 

This solution is only admissible if ${ p } \geq 3$, to ensure positivity of the argument of the square root. The local Lyapunov exponents, in this case are 
%With respect to p 
%\begin{equation}
%(\lambda_1,\lambda_2)=(-\frac32 + \frac32  \sqrt{9-\frac83  { p }}, -\frac32 - \frac32  \sqrt{9-\frac83  { p }}).
%\end{equation}
%With respect to y 
\begin{equation}
(\lambda_1,\lambda_2)=(-\frac32 + \frac32  \sqrt{1-\frac89 { y }^2},-\frac32 - \frac32  \sqrt{1-\frac89 { y }^2})
\end{equation}
and depending on the value of ${ p}$, the critical point falls under one of the following categories: 

\begin{itemize}
\item For  $3<{ p }<3.375$ or $0<{ y }<\frac{3\sqrt2}{4}$, the eigenvalues are real and $\lambda_1\neq\lambda_2 < 0$, so the system evolution, represented as the solid black line of figure\ref{fig:background}, leads towards the critical point that is a stable node in figure\ref{hyperstable}. 
\item For $ {p}>3.375$ or $y>\frac{3\sqrt2}{4}$, the eigenvalues are complex conjugates with negative real part and therefore the system - represented by the purple solid line in figure\ref{fig:background}- when perturbed, will spiral towards this stable focus, portrayed in figure\ref{hyperspiral}.

Shifting the hyperbolic field-space curvature function argument by a constant, bears no change in the dynamics. Hence, after redifining $f(\rho)\equiv\, e^{\left[{\rho}-{ \rho}(0) \right]}$, and using the linearised expression for $\rho-\rho(0)\approx -3N $ we see that in both the aforementioned cases the solutions evolve towards the hyperbolic solution as: 
\begin{equation}
\begin{split}
{\rho}=&{\rho}(0)-3N- \frac{c_1}{{ y } }\, e^{\lambda_1N}-\frac{c_2}{{ y }}\, e^{\lambda_2N}\\
{ \phi}=&{ \phi}(0) + \frac{c_1}{\mu_1} e^{\mu_1N}+ \frac{c_2}{\mu_2}e^{\mu_2N}\pm {y} \,e^{3N} 
\end{split}
\end{equation}
with  $(\mu_1,\mu_2)=(3+\lambda_1, 3+\lambda_2)$.

\item For $ { p }=3.375$, or $y=\frac{3\sqrt2}{4}$ there is a single eigenvalue, $\lambda=-\frac32$. Using the linearised solution for $\rho-\rho(0)$, again, we observe that the field follows the trajectory represented as the solid red line in figure\ref{fig:background}, towards the critical point that is an improper stable node in figure\ref{hyperimproper} and the hyperbolic solution takes the form: 
\begin{equation}
\begin{split}
{\rho}=&{\rho}(0)-3N- \frac23 \left[c_1 N + c_2\right] \, e^{-\frac32 N}\\
{ \phi}=&{ \phi}(0) + \frac{1}{{ y }}\left[\left(\frac32 N - 1\right)c_1 +\frac32 c_2 \right]e^{\frac32 N}\pm \frac{{y}}{3} \,e^{3N} .
\end{split}
\end{equation}

\end{itemize}
% critical point 
	\begin{figure}[!htb]
		\label{hyperphase}
	\subfloat[For  $3<{ p }<3.375$ the critical point is a stable node. 
	\label{hyperstable}]{%
	\includegraphics[width=0.28\textwidth]{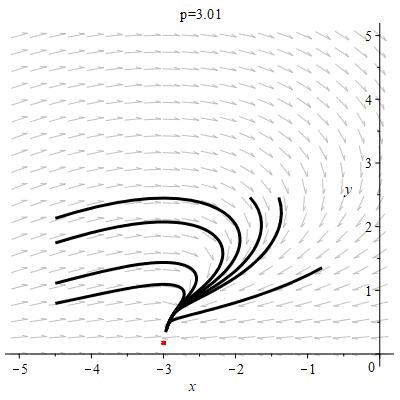}}
	\hfill
	\subfloat[For  ${ p }=3.375$ the critical point is improper stable node \label{hyperimproper}]{%
		\includegraphics[width=0.28\textwidth]{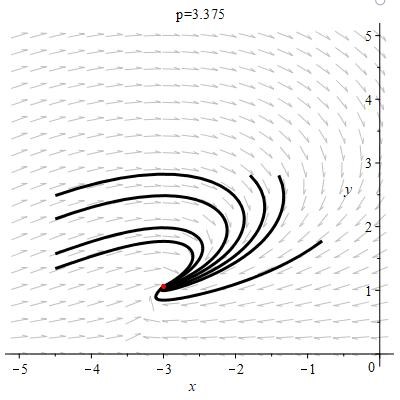}}
	\hfill
	\subfloat[For $ {p}>3.375$,the critical point is a stable focus. \label{hyperspiral}]{%
		\includegraphics[width=0.28\textwidth]{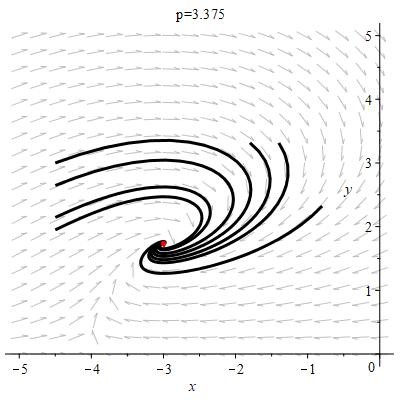}}
	\caption{Phase diagrams for the hyperbolic solution for various values of the potential gradient, ${ p }$.}
\end{figure}

%\begin{figure}[!htb]
%	\centering	
%	\begin{minipage}{0.4\textwidth}
%		\centering
%		\includegraphics[width=\linewidth]{figures/hyperbolic_phase.jpg}
%		\caption{\footnotesize{Phase diagram for the hyperbolic solution, with initial conditions perturbed from the critical point, for various values of $ { p }$}.} \label{hyperplot} 
%	\end{minipage}\hspace{0.5cm}
%	\begin{minipage}{0.4\textwidth}
%		\centering
%		\includegraphics[width=\linewidth]{figures/combo.jpg}
%		\caption{\footnotesize{Combined phase diagram, with initial conditions slightly perturbed away from the critical point, for various values of $ { p }$}.} \label{combo}
%	\end{minipage}\hspace{0.5cm}
%\end{figure}
\end{itemize}

%As $ { p }$ increases beyond $ { p }>3$, the system evolves towards the hyperbolic critical point \eqref{hypercrit}, maintaining $ { x }=-3$ and with $ { y }$ increasing as $ { y }\propto \sqrt{3 { p }-9}$. For various initial conditions and values of $ { p }$ the hyperbolic solution results in the following phase diagrams figure\ref{hyperphase}.

Finally, as is manifest form the linearised equations, the system close to the critical points evolves towards the minimum of the potential $\rho=0$, until the condition ${\rho}>1$ ceases to hold, that is when the approximation $f(\rho) = \sinh({ \rho}) \simeq e^{{\rho}}$ stops being valid.  
%\vspace{-0.3cm}

\section{The perturbations}

Having thoroughly explored the dynamics of homogeneous field configurations and their attractor scaling solutions, we now turn to the study of small amplitude inhomogeneities around backgrounds described by the hyperbolic solutions. Since it is well motivated that $M^2_H/M^2_p =\kappa^2 \ll 1$ \cite{Brown}, we will be dropping $\mathcal{O}(\epsilon)$ terms in the linearized perturbation equations - see (\ref{epsilon}) - from now on, effectively ignoring spacetime metric perturbations and their effect on the dynamics of field perturbations. 

\subsection{Linear perturbation analysis}
Assuming a general coordinatization of the field manifold, we can always split the field coordinates as 
\begin{equation}
\varphi^a(t,\mathbf{x}) = \bar{\varphi}^a(t) + q^a(t,\mathbf{x})
\end{equation} 
where the background $\bar{\varphi}^a(t)$ will be taken to follow the hyperbolic solution discussed above. The equations of motion for the perturbations $q^a$ then read
\cite{toms}
\begin{equation}
\label{perteqofmot1}
D_{\mu} D^{\mu}q^a  +  \mathcal{G}^{ab}q^d D_d  \,V,_b   +  \mathcal{R}^a_{bcd} \, q^d \partial_{\mu}\, \phi^b \,\partial^{\mu}\, \phi^c = 0.
\end{equation}
Projecting these equations along and orthogonal to the Killing (angular) direction, and using the results from \eqref{geometry}, leads to the following equations of motion for the geometrically normalised field perturbations  $(r,q)=(q^1,f(\rho) \, q^2)$,
\begin{equation}
\label{radial}
\ddot{ {r}} + 3 \, H \, \dot{ {r}}  - a^{-2} {\nabla_{\mathbf{x}}^2} {r} -2H^2 {y}^2 \,  {r} -2H {y} {\dot{q}} + 2H^2 \,  {x} \,   {y} \,  {q} =0 \\
\end{equation}
and 
\begin{equation}
\label{angular}
\ddot{ {q}} + 3 \, H \, \dot{ {q}}  - a^{-2}{\nabla_{\mathbf{x}}^2} {q} + 2\, H \,  {y} \, \dot{ {r}} + 3\, H^2  {q}\,  {p} - H^2\left( {x}^2 + {y}^2\right) {q} =0 ,
\end{equation}
%Their numerical solution will allow us to obtain the short-wavelength behaviour of the perturbation and eventually the associated noise that drives the quasi-classical long-wavelength behaviour.

The fluctuations of the field multiplet can be expressed  as a mode expansion in Fourier space 
\begin{equation}
\label{perturbationmode11}	
q^a(t,\mathbf{x})=\int \frac{d^3k}{(2\pi)^3} \,\,  \left[U^a_{\,\,\,\, I}(k,t)\, a^I_{\mathbf{k}}\, e^{i\, \mathbf{k}\cdot \mathbf{x}} +\, {U^{\star}}^a_{\,\,\,\, I}(k,t)\, a^{\dagger \, I}_{\mathbf{k}}\, e^{-i\, \mathbf{k}\cdot \mathbf{x}}\right]\,,
\end{equation}
noting that each multiplet member receives contributions from all creation/annihilation operators via the mode matrix $U^a{}_I(k,t)$. The normalisation is chosen such that
the creation/annihilation operators satisfy 
\begin{equation}
\left[a^I_{\mathbf{k}},{a_{\mathbf{k}'}^J}^{\dagger}\right]=\left(2\pi\right)^3 \delta^{IJ} \,  \delta(\mathbf{k}-\mathbf{k}').
\end{equation}  
The mode matrix involves 4 complex entries (obviously $N^2$ for an $N$-dimensional multiplet) but since the different fields are initially uncorrelated and evolve independently on small scales, there are only two independent modes corresponding to an initial condition where $U^a{}_I \propto \delta^a{}_I$. Later evolution, around and after horizon exit, mixes all the components of the matrix $U$. Below we will use two sets of modes, denoted by the index $I$. We choose these depending on whether the initial condition initially fluctuates along the adiabatic or entropic direction, which we now describe.

\subsection{Local Orthogonal Basis}

For the case of multiple fields, with background field velocity $\dot{\sigma}= \sqrt{G_{ab} \dot{\varphi}^a \, \dot{\varphi}^b}$, it is useful to define the adiabatic direction along the background field trajectory and isocurvature directions that are the normal to it \cite{Gordon, GrootNibbelink:2001qt}, defined by unit vectors $\hat{\sigma}$ and $\hat{s}$ with components 
\begin{equation}
(\hat{e}_{\sigma})^a=\frac{\dot{\phi}^a}{\dot{\sigma}} \hspace{3cm} (\hat{e}_{s} )^a= \frac{1}{\sqrt{\mathcal{G}}} \, \epsilon^{ac} \, {\sigma}^b \, \mathcal{G}_{bc}.
\end{equation}
This introduces a local orthogonal basis, related to the original coordinate basis through the zweibeins:
\begin{equation}
\begin{bmatrix} 
\hat{e}_{\sigma}  \\
\hat{e}_{s} \\
\end{bmatrix}=\begin{bmatrix} 
\frac{ {x}}{\dot{ {\sigma}}} & \frac{ {y}}{f \, \dot{ {\sigma}}}  \\
\frac{ {y}}{\dot{ {\sigma}}} & -\frac{ {x}}{f \,\dot{ {\sigma}}} \\
\end{bmatrix}
\cdot 
\begin{bmatrix} 
\hat{e}_{\rho}  \\
\hat{e}_{\phi} \\
\end{bmatrix},
\end{equation}
in which the covariant derivative is given by: 
\begin{equation}
\mathcal{D}_a V^b = \partial_a V^b + \slashed{\Gamma}^b_{ca} V^c
\end{equation}
with $\slashed{\Gamma}$ being the zweibein basis connection. The latter can be obtained in relation to the connection associated with the original coordinate basis, from \eqref{geometry}, $\Gamma^I_{JK}$, by: 
\begin{equation}
\slashed{\Gamma}^a_{b c}=e^K_{\ b} \, e^I_{\ c} \, e^{\   a}_{J} \Gamma^J_{IK} - e^{K}_{\ b} \,  e^{I}_{\ c} \, \partial_I \left(e_K^{\ a}\right)
\end{equation} 	
where now the capital indices are related to the coordinate basis and the lower-case ones to the zweibein.
In this basis, the only non-vanishing components 
%associated with the coordinate basis field space metric \eqref{fieldmetric}  
are:
\begin{equation}
\slashed{\Gamma}^1_{21}=\frac{ { y}}{ \, \dot{ { \sigma}  } } \hspace{1cm}  \textnormal{and} \hspace{1cm} 
\slashed{\Gamma}^2_{12}=\frac{  {x}   }{ \, \dot{     { \sigma}   }  } \ .
\end{equation}
Having established the relation between the two bases, we can transform equations \eqref{radial} and \eqref{angular} in the zweibein basis,
\begin{equation}
\begin{split}
\ddot{q}^{\sigma} + 3\, H \, \dot{q}^{\sigma} + \left(\frac{k}{a}\right)^2 q^{\sigma} + 6\, H^2 \, {y}\,  q^{s} + 2 \, H\,   {y} \, \dot{q}^{s}=0& \\
\ddot{q}^s + 3\, H \, \dot{q}^s + \left(\frac{k}{a}\right)^2 q^s -2\, H^2\,  {y}^2 q^s -2 \, H\,   {y} \, \dot{q}^{\sigma}=0&
\end{split}
\end{equation}
%or, using $\eta=-\frac{k}{\alpha\, H}$
%\begin{equation}
%\begin{split}
%\label{eqofmoteta}
%\partial_{\eta}^2\, q^{\sigma}-2\, \frac{1}{\eta} \partial_{\eta} q^{\sigma} + q^{\sigma} +6 \, \frac{\tilde{y}}{\eta^2} \, q^{\sigma} -2\, \frac{\tilde{y}}{\eta} \, \partial_{\eta} q^{s}=0& \\
%\partial_{\eta}^2\, q^{s} -2\, \frac{1}{\eta} \partial_{\eta} q^{s}+q^s -2\, \frac{\tilde{y}^2}{\eta^2} \, q^s +2\, \frac{\tilde{y}}{\eta} \, \partial_{\eta} q^{\sigma}=0& .
%\end{split}
%\end{equation}
%In the limit $\tilde{y}\rightarrow 0$, the effect of the field-space curvature vanishes and the fields decouple, reproducing the standard single field equations of motion in each direction, which assume a set of normalised solutions, in four spacetime dimensions \cite{Miao},
%
%\begin{equation}
%\label{singlefieldmodes}
%u_k(t)=\alpha^{-3/2} \frac{\sqrt{\pi}}{2} \mathcal{H}^{(1)}_{\nu}\left(\frac{-k}{\alpha H} \right) {H}^{-1/2} \, ,
%\end{equation}
%where $\mathcal{H}_{\nu}$ is the Hankel function, with subscript $\nu=\sqrt{9/4-m^2/H^2}$. 
%For light scalar fields, $\nu=3/2$ and the mode function takes the much simpler form 
%\begin{equation}
%\label{Henkel}
%u_k=  \frac{ H  }{ \sqrt{k^3}} \  \frac{\sqrt{\pi} }{2}\,  (-\eta)^{\frac32} \, H_{\frac32 } \left(
%%-
%\eta \right) .
%\end{equation}
where we used \eqref{perturbationmode11}, $q^{\sigma}(\mathbf{k},t) = U^\sigma{}_I(k,t)\, a^I_{\mathbf{k}}$ and similarly for $q^s$. Introducing $z=Ht+\ln(H/k)$, the equation of motion for the modes becomes:
\begin{equation}
\begin{split}
\label{eqofmotz}
\partial_{z}^2\, q^{\sigma} + 3 \,  \partial_{z} q^{\sigma} + e^{-2z} q^{\sigma} + 6 \,  {y} \, q^{s} +2\,  {y} \, \partial_{z} q^{s}=0& \\
\partial_{z}^2\, q^{s} + 3 \, \partial_{z} q^s +e^{-2z} q^s -2\,  {y}^2 \, q^s - 2 \,  {y} \, \partial_{z} q^{\sigma}=0& .
\end{split}
\end{equation}
%Note that our time variable $z$ is simply the number of e-folds up to a $k$-dependent shift and that all modes experience the same time evolution but only shifted in time, with $z=0$ denoting horizon exit for each mode. { \color{red} In the limit of radial motion, $ {y}\rightarrow 0$, the effect of the field-space curvature vanishes and the fields decouple, hence reproducing the one dimensional equations of motion, the solution of which are the positive frequency k-modes for a massless minimally coupled scalar field on a de Sitter background, defining the Euclidean or Bunch–Davies vacuum state \cite{Bunch} }
%and assume a set of normalised solutions, in four spacetime dimensions \cite{Miao},

Note that our time variable z is simply the number of e-folds up to a k-dependent shift
and that all modes experience the same time evolution but only shifted in time, with $ z = 0$
denoting horizon exit for each mode. Initial conditions are set as usual when each mode is
deep inside the Hubble radius and the fields are effectively decoupled and massless. Assuming
that the state is described by the Bunch-Davies vacuum the relevant mode functions are given
	by \cite{Bunch, Miao},
\begin{equation}
\label{Henkel}
u_k=  \frac{ H  }{ \sqrt{k^3}} \  \frac{\sqrt{\pi} }{2}\,  (-\eta)^{\frac32} \, H_{3/2 } (k  \eta),
\end{equation}
where $\eta=-e^{-z}$,

\subsection{Hyperinflation Perturbation Power Spectra}
\label{icsection}
\begin{figure}[H]
	\centering	
	\begin{minipage}{0.45\textwidth} 
		\centering
		\includegraphics[width=\linewidth]{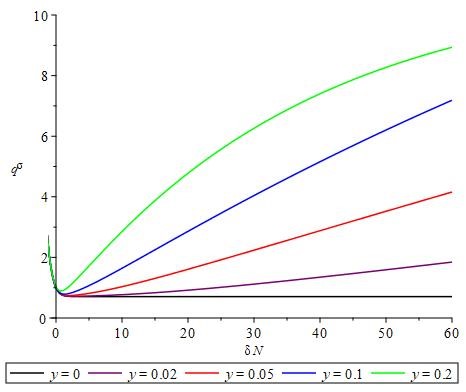}
	\end{minipage}\hspace{1cm}
	\begin{minipage}{0.45\textwidth} 
		\centering
		\includegraphics[width=\linewidth]{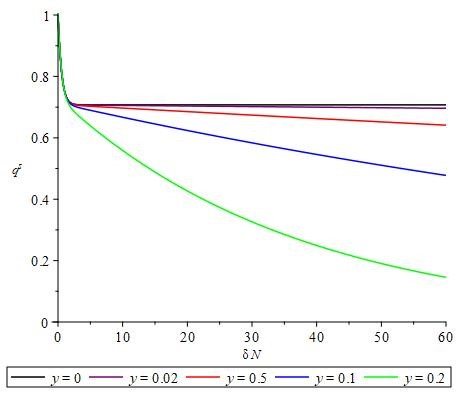}
	\end{minipage}
	\caption{\footnotesize Time evolution of the adiabatic (left) and entropic (right) modes, for various values of $ { y }$. The horizontal semi-axes' origin corresponds to the time of horizon exit for a mode with wavenumber $k$ that is followed for further 60 e-folds. Increasing values $y$, or the potential slope $p$, lead to a faster increase of the adiabatic perturbation and decrease of the isocurvature one.        
	}\label{xisigma}
\end{figure}

Deep inside the horizon the spatial gradient term of equations (\ref{eqofmotz}) dominates and therefore the modes are decoupled and evolve independently, unaffected by the field-space curvature induced mixing. As discussed above, we take our two sets of modes to be defined as $z\to-\infty$ by
\begin{equation}
U^a{}_I \to u_k \, \delta^a{}_I
\end{equation}
where now $a=(\sigma,s)$ denotes the adiabatic and isocurvature directions.
%\begin{align}
%&U^\sigma{}_1\to u_k&&U^s{}_1\to 0\\
%&U^\sigma{}_2\to 0&&U^s{}_2\to u_k,
%\end{align}
Defining rescaled modes $\mathcal{U}^a_{\ I}$
\begin{equation}
\label{xinorm}
U^a_{\ I}\equiv \frac{H}{\sqrt{k^3}} \, \mathcal{U}^a_{\ I}.
\end{equation}
the field power spectra can then be written as
\begin{equation}
\begin{split}
\mathcal{P}^{ab}(k,t) &= \frac{k^3}{2\pi^2} \int d^3x \,e^{-i\mathbf{k}\cdot \mathbf{x}} \langle q^a(t,\mathbf{0}) q^b(t,\mathbf{x})\rangle \\
&= \frac{H^2}{2\pi^2}\left(\sum_I \mathcal{U}^{a\star}_I \,  \mathcal{U}^b_I\right).
\end{split}
\end{equation}
To obtain the modes we solve equations (\ref{eqofmotz}) twice, once with $q^a = U^a{}_1$ and initial conditions, $(U^\sigma{}_1\to u_k\,,U^s{}_1\to 0)$, and once with $q^a = U^a{}_2$ and initial conditions $(U^\sigma{}_2\to 0\,,U^s{}_1\to u_k)$. In practise, the calculation starts at a sufficiently negative $z$ such that the two equations are effectively decoupled. and we solve them for $z = 60$ e-folds after horizon exit, defining the longest mode of interest. The evolution of the modes can be seen in figure $\ref{xisigma}$. 

\begin{figure}[H]
	\centering	
	\begin{minipage}{0.6\textwidth}
		\centering
		\includegraphics[width=\linewidth]{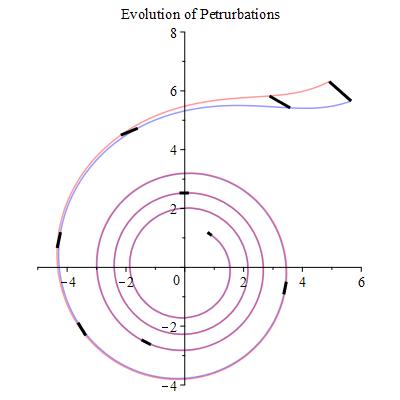}
		\caption{\footnotesize{Evolution of the linearised field perturbations with parameter ${ p}=3.01$ or ${ y}\approx 0.17$. The adiabatic (tangent to the trajectory) perturbations approach a small constant value. It is worth noting that the apparent shrinkage is due to the choice of coordinates. \label{fig:LW} }}
	\end{minipage}\hspace{0.5cm} 
\end{figure}

To corroborate the superhorizon behaviour of the linearized field perturbations as obtained by the numerical solution presented above, we have also followed the evolution of two ``separate universes'' that start off at slightly displaced points in field space. Figure \ref{fig:LW} shows one such configuration, where two solutions of the background equations with neighbouring initial conditions are allowing to evolve, while depicting their difference at equal times in field space. In this particular example we see that an initially dominant entropic perturbation feeds into an initially subdominant adiabatic one and decays while the adiabatic perturbation, corresponding to a time shift of a single background solution, eventually evolves towards a small, constant value. (However, this only occurs after an observationally inaccessible number of e-folds.)

\begin{figure}[H]
	\centering	
	\begin{minipage}{0.45\textwidth}
		\centering
		\includegraphics[width=\linewidth]{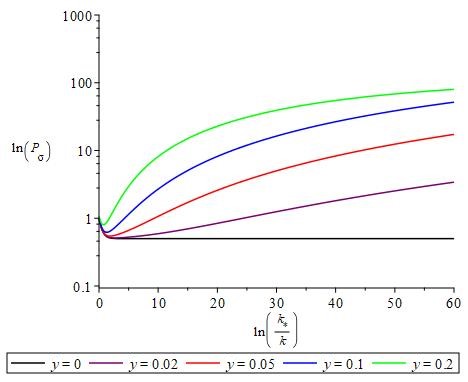}
		%	\caption{\footnotesize Time evolution of the adiabatic Power Spectrum, $\mathcal{P}_{\sigma}(z)$ for various values of $ { y }$}
	\end{minipage}\hspace{0.5cm}
	\begin{minipage}{0.45\textwidth}
		\centering
		\includegraphics[width=\linewidth]{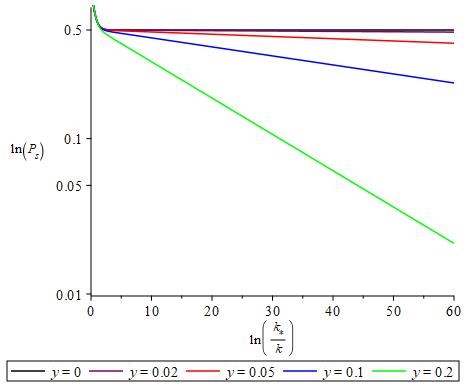}
		%	\caption{\footnotesize Time evolution of the entropic Power Spectrum, $\mathcal{P}_{s}(z)$ for various values of $ { y }$}
	\end{minipage}	\caption{\footnotesize The adiabatic $\mathcal{P}_{\sigma}$ and entropic Power Spectra, $\mathcal{P}_{s}$ at time $t$ for various values of $ { y }$, where  $k^*$ is the mode exiting the horizon at that time and  $\ln{(k_*/k)}=60$ represents the wavenumber corresponding to the scale of the Universe today. }
	\label{powerspectrum}
\end{figure}

The power spectra are easy to obtain as a function of $k$ since all $k$ modes follow the same evolution, only shifted in time by $t \rightarrow t+H^{-1}\ln(H/k)$, see the definition of $z$ above (\ref{eqofmotz}). This is of course a consequence of the hierarchy of the energy scale  of the scalar field background $M_H \ll M_P$, resulting in $H$ being treated as a constant. In general, if the value of H changed, each mode would cross the horizon at a different amplitude and their subsequent temporal evolution after horizon crossing would vary, rendering this identification inaccurate. We plot the power spectra for different hyperbolic scaling attractors, characterized by different values of $y$, in figure \ref{powerspectrum}.

\begin{figure}[H]
	\centering	
	\centering
	\includegraphics[width=0.6\linewidth]{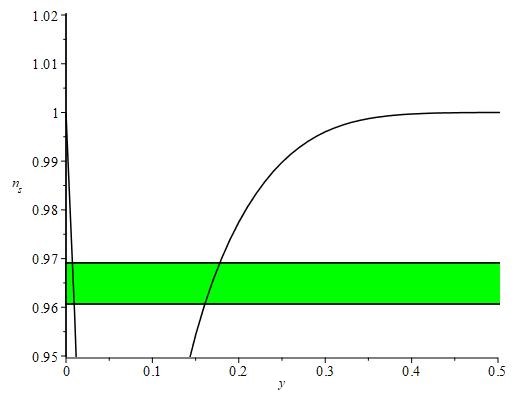}
	\caption{The Adiabatic Spectral Index, $n_s$ as a function on $ { y }$. The green band corresponds to observationally permitted values \cite{Akrami:2018odb}.}
	\label{spectral}
\end{figure}

The adiabatic spectral index 
\begin{equation}
n^{ad}_s=1+\frac{d\ln\left[ 
	P^{\sigma\sigma} \right] }{d(\ln k)}
\end{equation} 
is plotted in figure \ref{spectral}, along with the $68\%$ confidence range for its value from the Planck collaboration \cite{Akrami:2018odb} (green band $n_s =  0.9649 \pm 0.0042$ ).  As we see, the measured value places very tight bounds on the permissible values for $y$, to two narrow ranges: $0.74 \times\, 10^{-2}\leq {y}\leq 0.94 \,\times \,  10^{-2}$ and  $0.1603 \leq {y}\leq 0.1774$, severely constraining realistic  hyperinflation models to those with exponential potential parameters  in the ranges $0.54 \times 10^{-4}\leq p-3\leq0.88\times 10^{-4} $ and $0.025\leq p-3\leq 0.031$ .

Finally, we look at the tensor-to-scalar ratio. We have at horizon crossing, 
\begin{equation}
r=\frac{{\cal P}_t(k)}{{\cal P}_s(k)}
%{4 \kappa^2} \frac{{\varphi^{\prime}}_{*}^2}{\, U^2}
=\frac{8 \epsilon}{U^2}=\frac{4\kappa^2}{U^2} \,\left[9+ { y}^2 \right].
\end{equation}
where $U=\sum_I \mathcal{U}^{\sigma\star}_I \,  \mathcal{U}^\sigma_I$ and we have only taken the adiabatic perturbations into account as the contribution of the isocurvature ones is negligible. The  Planck 2018 collaboration \cite{Akrami:2018odb} set the observational bound for the tensor-to-scalar ratio to $r<0.10$. Hence, taking into account that 
%(if chosen such that the perturbation solutions are  is normalised by 1/L, see note before eq.3.11 )
$U^2 < 50$ (see figure\ref{powerspectrum}) and using \eqref{hypercrit}, we obtain bounds for the parameter of the exponential potential:
\begin{equation}
\label{kappapi}
\kappa^2 {p}<\frac{5}{12}. 
\end{equation}
This, in turn, results in limits for the permitted values for the Hubble parameter $\epsilon$:
\begin{equation}
\begin{split}\label{epsilon1}
\epsilon
&=\frac12\kappa^2\left[ {x}^2+ {y}^2 \right]\\
&=\frac32 \kappa^2\, {p}\\
&<5/8.
\end{split}
\end{equation}
Equation \eqref{epsilon1} shows that our original ansatz of small $\epsilon$ is observationally consistent. %This implies that this type of exponential potential hyperinflation models, which is the simplest type of those that are not observationally excluded (as opposed to those with a power-law potential \cite{Mizuno}), cannot elevate (hyper)inflation out of the Swampland, which posits that $\epsilon \approx \mathcal{O}(1)$, 
% unless the requirement of the conjecture is relaxed to $\epsilon \approx \mathcal{O}(10^{-1})$ as originally argued in \cite{Bjorkmo}. This conclusion can be made more robust by extending our analysis beyond the small $\epsilon$ regime. 

It is note-worthy that the definitions of the Hubble slow-roll parameter $\epsilon\equiv-\dot{ H}/H^2$, given by \eqref{epsilon}, and that of the potential slow-roll parameter % constrained by the aforementioned observations, and that of 
\begin{equation}
\label{epsilonV}
\epsilon_V\equiv \frac12 \kappa^{-2}\  \frac{V^a\, V_a}{V^2} = \frac12 \kappa^2 \ p^a \,\mathcal{G}_{ab} \,  p^b, 
\end{equation} 
%which according to the deSitter Swampland Conjecture should be of order $\mathcal O(1)$, 
are inequivalent in multi-field models and can differ substantially \cite{Achucarro:2018vey}. Equation \eqref{epsilonV} is covariant, hence, working in the convenient $(x,y)$ coordinate system defined underneath \eqref{epsilonprime}, one readily obtains 
\begin{equation}
\label{last}
\epsilon_V=\frac12 \kappa^2 p^2 \frac{x^2}{x^2+y^2}.
\end{equation}
%
%The aforementioned observational constraints, namely $p_{max}=3.031$ and $y^2=\mathcal{O}(10^{-3})$ place very tight bounds on the evolution of the system very close to the gradient solution $p=3$. It is unsurprising, hence, that in the simple exponential potential hyperinflation model under consideration here, $\epsilon_V$ does not substantially vary from $\epsilon$, 

Due to the characteristics of our particular model, namely that  we consider scaling solutions % $\epsilon_H^{\prime}=0$, resulting in \eqref{hypercrit} in the $p>3$ regime, 
and use a potential that respects the field-space isometries, resulting in \eqref{last} receiving contributions only from the radial direction, it is evident that the two slow-roll parameters coincide,

\begin{equation}
\epsilon_V=\frac12 \, \kappa^2 \, p^2 \frac{9}{3p}=\epsilon_H < 5/8.
\end{equation}
Finally, this indicates that this type of exponential potential hyperinflation models, which is the simplest type of those that are not observationally excluded (as opposed to those with a power-law potential \cite{Mizuno}), cannot elevate (hyper)inflation out of the Swampland unless the requirement of the conjecture is relaxed to $\epsilon_V \approx \mathcal{O}(10^{-1})$ as originally argued in \cite{Bjorkmo}. This conclusion can be made more robust by extending our analysis beyond
the small $\epsilon_H$ regime.

\section{Summary and Motivation for Further Work }

In this work we have reviewed Hyperinflation in an exponential potential and constrained its slope to be in agreement with observational results. Adopting a hierarchy of scales $M_H<\phi<M_{P}$ results in the slow-roll parameter $\epsilon$ being negligible while its time derivative vanishes in the attractor Hyperinflationary solution. This results in the equations of motion for the background field taking the form of an autonomous system. After solving the linearised equations, we have manifestly shown the dynamical patterns that describe the behaviour of the system for various values of the exponential parameter $p$. Furthermore, the linear stability of the background has been manifestly demonstrated and two different regimes have been distinguished: For $p<3$, the gradient solution is a stable attractor and for $p>3$, where the gradient solution is a reppeler, the hyperinflation solution dominates.
Furthermore, by introducing a local orthogonal basis, we obtained the equations of motion of the adiabtic and isocurvature perturbations and noted their highly coupled behaviour that is due to the field space curvature. We proceeded to solve these numerically and described the dynamics of the perturbations, noting their behaviour after horison crossing: The adiabatic perturbations grow with the parameter $y$ associated with the field-space curvature and the entropic ones rapidly decay. 

Finally, after producing the adiabatic and entropic power spectra, we compared the resulting adiabatic spectral index and tensor to scalar ratio with recent observational values, placing tight bounds on the permissible values of the exponential potential parameter $p$, also setting an upper limit for the $\epsilon$  parameter:  $\epsilon \lesssim \mathcal{O}(10^{-1})$. 

Hyperinflation is a regime that could shine light on intriguing aspects of inflationary cosmology. It has been shown \cite{Christodoulidis:2019mkj} that this model is a specific case of side-tracked inflation, namely the attractor phase emerging from geometrical destabilisation \cite{Renaux-Petel:2015mga}. 
Brown \cite{Brown}, in his original paper, notes the equilateral non-Gaussianities that the model can give rise to and substantial work has been carried out for the computation of such non-Gaussianities in a general class of multi-field models with a curved field-space \cite{Garcia-Saenz,Garcia-Saenz:2018vqf,Fumagalli:2019noh, Bjorkmo:2019qno, Ferreira:2020qkf} . It would be of particular interest to the authors to investigate such higher-point functions in the large-wavelength limit using the technology of stochastic inflation. 
%The high degree of non-linearity in the equations of motion for the perturbations, especially when the spacetime deviates from deSitter and a backreaction on the expansion rate $H=H(\phi)$ is taken into account. 
We leave the derivation of the relevant stochastic equations and the computation of the arising non-Gaussianity for future work.

%\appendix
%\clearpage
\section*{Acknowledgements}
\noindent I. G. M. is supported in part by the STFC grant [ST/T000708/1].\\
M.B is supported by a Newcastle University Studentship.

\end{document}